\documentclass[useAMS,usenatbib]{mn2e}
\include{graphicx}
\title[Detection of Earth and Mars Trojans by Gaia]{Predictions for the Detection of Earth and Mars Trojan Asteroids by the Gaia satellite}

\author[M. Todd, P. Tanga, D. M. Coward and M. G. Zadnik]{M. Todd$^{1}$\thanks{E-mail:
michael.todd@icrar.org (MT)}, P. Tanga$^{2}$, D. M. Coward$^{3}$ and M. G. Zadnik$^{1}$\\
$^{1}$Department of Imaging and Applied Physics, Bldg 301, Curtin University, Kent St, Bentley, WA 6102, Australia\\
$^{2}$Laboratoire Lagrange, UMR7293, Universit\'{e} de Nice Sophia-Antipolis, CNRS, Observatoire de la C\^{o}te d'Azur, France\\
$^{3}$School of Physics, M013, The University of Western Australia, 35 Stirling Hwy, Crawley, WA 6009, Australia}

\begin{document}

\date{Accepted 2013 November 13.  Received 2013 November 11; in original form 2013 September 26}

\pagerange{\pageref{firstpage}--\pageref{lastpage}} \pubyear{}

\maketitle

\label{firstpage}

\begin{abstract}
The European Space Agency \textit{Gaia} satellite, planned for launch in late 2013, will perform systematic astrometric observations of the whole sky over a five year period. 
During this mission many thousands of Solar System Objects down to magnitude $V=20$ will be observed including Near-Earth Asteroids and objects at Solar elongations as low as $45\degr$, which are difficult to observe with ground-based telescopes. 
We simulated the detection of Trojan asteroids in the orbits of Earth and Mars by \textit{Gaia}. 
We find that \textit{Gaia} will not detect the Earth Trojan 2010~TK$_7$ although it will detect any Earth Trojans with diameters larger than 600~m. 
We also find that \textit{Gaia} will detect the currently known Mars Trojans and could discover more than 100 new Mars Trojans as small as 400~m in diameter. 
The results of the \textit{Gaia} mission will test the predictions about the Mars Trojan asteroid population and lead to greater understanding about the evolution of the Solar System.

\end{abstract}


\section{Introduction}

\setcounter{footnote}{3}
The European Space Agency (ESA) \textit{Gaia} satellite (http://gaia.esa.int) will be launched in late 2013. \textit{Gaia} will be located near the Earth's L2 Lagrangian point and will survey the whole sky down to magnitude $V=20$ \citep{2007EM&P..101...97M}. In the course of its mission several tens of thousands of asteroids will also be repeatedly observed. Observations of Main Belt asteroids and Near Earth Objects by \textit{Gaia} have been looked at extensively during the past several years in papers such as \citet{2007EM&P..101...97M} and \citet{2012P&SS...73....5T}. However, the sky coverage will also include the regions in the orbits of Earth and Mars where Trojan asteroids are likely to exist \citep{2012P&SS...73...39T, 2012MNRAS.420L..28T, 2012MNRAS.424..372T}. We anticipate obtaining results from \textit{Gaia} that will include the discovery of new Trojan asteroids.

Trojan asteroids occupy the L4 and L5 Lagrangian regions in a planet's orbit about the Sun. These stable regions exist $60\degr$ ahead (L4) and behind (L5) the planet in its orbit and appear to be stable on long time-scales in the N-body case of the Solar System \citep{1999CeMDA..73..117P, 2005Icar..175..397S}. Studying the Trojans provides insight into the evolution of the Solar System. Several thousand Trojans have been found in the orbit of Jupiter, however, among the terrestrial planets only Earth and Mars are known to have Trojans. 

Earth has only one known Trojan (2010 TK$_7$), discovered through examination of data from the \textit{WISE} satellite \citep{2011Natur.475..481C}, while Mars presently has five known Trojans (Eureka, 1998~VF$_{31}$, 1999~UJ$_7$, 2007~NS$_2$ and 2001~DH$_{47}$) listed by the Minor Planet Center. Previous modelling \citep{1999ApJ...517L..63T, 2000MNRAS.319...63T, 2000MNRAS.319...80E} suggests that there may be ten times this number of Mars Trojans with diameter larger than 1~km, and that there may be several hundred with diameter larger than 100~m.

Studies of three of the Mars Trojans (Eureka, 1998~VF$_{31}$ and 1999~UJ$_7$) by \citet{2003Icar..165..349R, 2007Icar..192..434R} show that these Trojans appear to have separate origins from one another. \citet{2007Icar..192..434R} suggested that the simplest explanation for the origins of Eureka and 1998~VF$_{31}$ is that they formed separately in other parts of the inner Solar System as part of larger bodies and that fragments were trapped in the 1:1 resonance with Mars roughly 3.9~Ga ago, thus they are most likely to be long-term residents rather than natives. 

Previous work \citep{2012MNRAS.420L..28T, 2012MNRAS.424..372T} considered strategies for discovering Earth and Mars Trojans using ground-based telescopes as well as considering the potential benefits of the \textit{Gaia} mission to the discovery of additional Trojans. We subsequently ran simulations to model detection characteristics and detection limits for Earth and Mars Trojans with respect to \textit{Gaia} using a distribution of test particles with a uniform size of 1~km and having Trojan-like orbits \citep{2012arXiv1212.0268T} and found that these would typically be bright enough (mag. $V<20$) but that the relative sky motion may be problematic depending on the observing geometry. In this paper we use the known Trojans to model the detections of additional Earth and Mars Trojans by \textit{Gaia}. In our simulations we also included the asteroids 2011~SC$_{191}$, 2011~SL$_{25}$ and 2011~UN$_{63}$ which have been proposed as Mars Trojans by \citet{2013MNRAS.432L..31D}. We refer to these proposed Trojans hereafter as Trojans for simplicity.

\section{Model}

For our simulations we used the known orbital elements for the Earth and Mars Trojans. One challenge that we faced is that we cannot be certain when \textit{Gaia} will begin observations. We do not yet know the starting epoch of the observations nor can we predict \textit{Gaia}'s orientation at that epoch. Thus it is not possible to simulate the real detection sequence. To compensate for this in assessing whether \textit{Gaia} will detect the known Trojans, we cloned the elements to provide 100 test bodies for each of these Trojans. We then modified the cloned elements to compensate for the uncertainty of the starting conditions. 
Although these orbits would not be stable over a long period the perturbations are trivial over the five years duration of the \textit{Gaia} mission simulation.
The modifications are described in \S\ref{section:earthtrojans} and \S\ref{section:marstrojans} for Earth Trojans and Mars Trojans respectively. 

\subsection{Earth Trojans}\label{section:earthtrojans}

\begin{figure}
\includegraphics{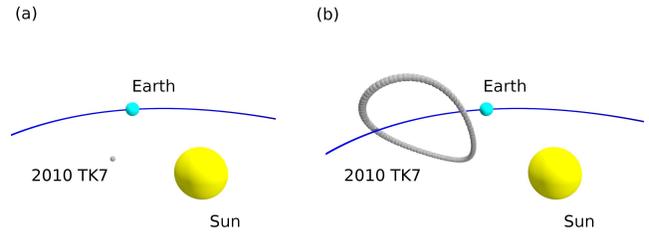}
\caption{\label{fig:figure1} (a) The starting position of Earth's Trojan asteroid, 2010~TK$_7$, and (b) the starting positions of the clones. }
\end{figure}

The Earth Trojan 2010~TK$_7$ is co-orbital with Earth. It resides in the L4 (leading) Lagrangian region and so travels around the Sun ahead of Earth in its orbit. Because it is co-orbital with Earth and hence also with \textit{Gaia}, it was slightly more complex to correct for uncertainties in the starting conditions for the observations by \textit{Gaia} than the Mars Trojans. Starting from the initial position of 2010~TK$_7$ (Fig.~\ref{fig:figure1}a), we created 100 objects by replicating the orbital elements (Table~\ref{tab:table1}) but translating both the ascending node and the mean anomaly. Translating the ascending node spaced the clones of 2010~TK$_7$ at intervals through $360\degr$ such that each clone started from the same ecliptic latitude, effectively rotating the orbit orientation. However, this correction on its own produced a set of false positives because it did not take into account that the 2010~TK$_7$ is co-orbital with Earth. Adjusting the mean anomaly in conjunction with the ascending node provided the necessary adjustment to create a set of clones of 2010~TK$_7$ which were co-orbital with Earth. Since \textit{Gaia}, Earth and 2010~TK$_7$ are co-orbital, by visualising the result in a \textit{Gaia}-centric frame we see the apparent motion of 2010~TK$_7$ (Fig.~\ref{fig:figure1}b).
We consider that the drift in position over time as it librates about the L4 point will not significantly affect the results of this simulation given the length of the \textit{Gaia} mission (5~years) since the period of libration is 395~years \citep{2011Natur.475..481C}.

\begin{table}
 \centering
\caption{Orbital parameters for the Earth Trojan asteroid 2010~TK$_7$ (semi-major axis $a$, eccentricity $e$, inclination $i$, ascending node $\Omega$, argument of perihelion $\omega$). \label{tab:table1}}
\begin{tabular}{lcccrr}
\hline 
Asteroid                 & $a$        & $e$        & $i$       & \multicolumn{1}{c}{$\Omega$}       & \multicolumn{1}{c}{$\omega$}	      \tabularnewline
\hline 
2010~TK$_7$              & 1.000248   & 0.19079    & 20\fdg 882  & 96\fdg 528  &  45\fdg 852   \tabularnewline
\hline 
\end{tabular}
\end{table}

\subsection{Mars Trojans}\label{section:marstrojans}

\begin{figure}
\includegraphics{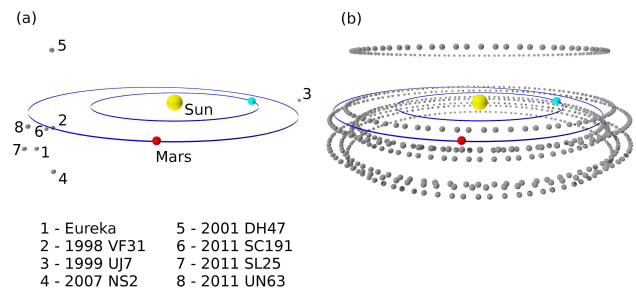}
\caption{\label{fig:figure2} (a) The starting position of the Mars Trojan asteroids, (b) the starting positions of the clones with the ascending node adjusted. }
\end{figure}

The Mars Trojans are co-orbital with Mars. The observing geometries for an Earth-based instrument, and for \textit{Gaia}, change with time due to the different orbital periods of Earth and Mars (and hence the Mars Trojans). 
For each of the Mars Trojans we took the initial positions (Fig.~\ref{fig:figure2}a) and replicated the orbital elements (Table~\ref{tab:table2}) to create 100 instances of each Trojan. Similarly to the simulation for the Earth Trojan 2010~TK$_7$ we translated the ascending node of each group of clones to space them at intervals through $360\degr$ (Fig.~\ref{fig:figure2}b), thereby rotating the orientation of their orbits. However, unlike the Earth Trojan simulation we did not translate the mean anomaly of the clones as they are not co-orbital with Earth.

\begin{table*}
 \centering
 \begin{minipage}{101mm}
\caption{Orbital parameters for the Mars Trojan asteroids (semi-major axis $a$, eccentricity $e$, inclination $i$, ascending node $\Omega$, argument of perihelion $\omega$). \label{tab:table2}}
\begin{tabular}{lcccrr}
\hline 
Asteroid                 & $a$        & $e$        & $i$       & \multicolumn{1}{c}{$\Omega$}       & \multicolumn{1}{c}{$\omega$}	      \tabularnewline
\hline 
Eureka            & 1.5235160  & 0.0646937  & 20\fdg 28289  & 245\fdg 06953  &  95\fdg 41530   \tabularnewline
1998~VF$_{31}$  & 1.5241558  & 0.1003765  & 31\fdg 29733  & 221\fdg 33011  & 310\fdg 55717   \tabularnewline
1999~UJ$_7$     & 1.5245294  & 0.0392950  & 16\fdg 74860  & 347\fdg 38708  &  48\fdg 35302   \tabularnewline
2007~NS$_2$     & 1.5237155  & 0.0540248  & 18\fdg 62167  & 282\fdg 49888  & 176\fdg 93881   \tabularnewline
2001~DH$_{47}$           & 1.5237644  & 0.0347226  & 24\fdg 39962  & 147\fdg 42910  &  17\fdg 59104   \tabularnewline
2011~SC$_{191}$          & 1.5238022  & 0.0441686  & 18\fdg 74444  &   5\fdg 79590  & 196\fdg 34215   \tabularnewline
2011~SL$_{25}$           & 1.5239213  & 0.1144574  & 21\fdg 49666  &   9\fdg 42731  &  53\fdg 29126   \tabularnewline
2011~UN$_{63}$           & 1.5237402  & 0.0645860  & 20\fdg 36251  & 223\fdg 57138  & 165\fdg 29725   \tabularnewline
\hline 
\end{tabular}
\end{minipage}
\end{table*}

\section{Results}

\textit{Gaia} will not directly produce images of each source. For these objects, since the brightness is typically at magnitudes below $V=16$, a window 6~x~6 pixels centred on the source will be assigned when an object is detected. However if the along-scan motion is greater than 3.5~mas~s$^{-1}$ the object will drift out of this window, resulting in a loss of signal. The median absolute along-scan velocity for Main Belt asteroids is 7~mas~s$^{-1}$ \citep{2012P&SS...73....5T} which is a drift of about half a pixel during a single CCD crossing.  A similar case holds for the across-scan motion although the pixel dimensions (59~mas x 177~mas) mean that a drift across-scan up to 10.5~mas~s$^{-1}$ will result in the object remaining within the window by the time the object has crossed the CCD array. The absolute across-scan velocity for Main Belt asteroids is typically less than 15~mas~s$^{-1}$ \citep{2012P&SS...73....5T}. Previous results for the detection of Earth and Mars Trojans by \textit{Gaia} \citep{2012arXiv1212.0268T} found that the along-scan and across-scan velocities for Earth and Mars Trojans are much higher than Main Belt asteroids. This has the implication that they will frequently drift out of the window during a crossing, resulting in signal degradation.

\subsection{Earth Trojans}

\begin{figure}
\includegraphics{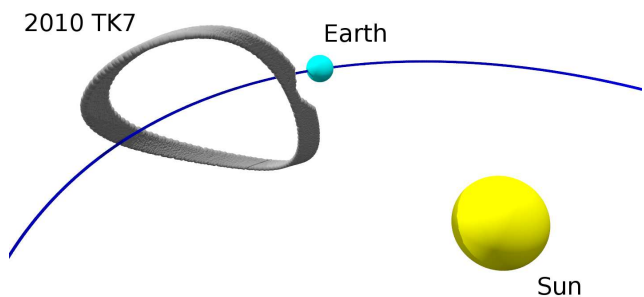}
\caption{\label{fig:figure3} Plotting the positions of the Earth Trojan asteroid 2010~TK$_7$ clones relative to \textit{Gaia}'s orientation shows that most of the orbits are within the region observable by \textit{Gaia}. 
There is a small gap where the orbits pass through the unobservable region at Solar elongations nearer than $45\degr$ from the Sun. }
\end{figure}

After replicating the orbital parameters for 2010~TK$_7$ (Table~\ref{tab:table1}) and simulating observations by \textit{Gaia} we plotted the observed positions of the clones relative to Earth (and \textit{Gaia}) (Fig.~\ref{fig:figure3}). 
The Solar elongations of these positions range between $44\fdg 6$ and $100\fdg 4$ (Table~\ref{tab:table3}) and the distance varies between 0.18~AU and 0.87~AU.

The apparent magnitude of an Earth Trojan with a diameter of 1~km ranges between $V = 17.9$ and $V = 19.5$, assuming an albedo of 0.20 \citep{2012MNRAS.420L..28T}. This variation depends on both phase angle and distance, with distance having the greater effect. Our simulation results for the Earth Trojan 2010~TK$_7$ indicate that it will not be detected by \textit{Gaia} in spite of its proximity to Earth. 
The Earth Trojan 2010~TK$_7$ has a diameter of only $\sim 300$~m \citep{2011Natur.475..481C}. The consequence of this smaller size is that its brightness is correspondingly reduced, ranging from $V = 20.9$ to $V = 22.8$ (Table~\ref{tab:table3}). As a result 2010~TK$_7$ is always too faint to be detected by \textit{Gaia}.

\begin{table}
 \centering
\caption{Observation statistics for Earth Trojan 2010~TK$_7$. Note the apparent magnitude of 2010~TK$_7$ is below \textit{Gaia}'s detection threshold. \label{tab:table3}}
\begin{tabular}{lccc}
\hline 
Asteroid                 & Distance from  & Solar    & Magnitude  \tabularnewline
                         & Earth (AU)   & elongation    & $(V)$      \tabularnewline
\hline 
2010~TK$_7$              & ${0.18 - 0.87}$       & ${44\fdg 6 - 100\fdg 4}$      & ${20.9 - 22.8}$       \tabularnewline
\hline 
\end{tabular}
\end{table}

\begin{figure}
\includegraphics{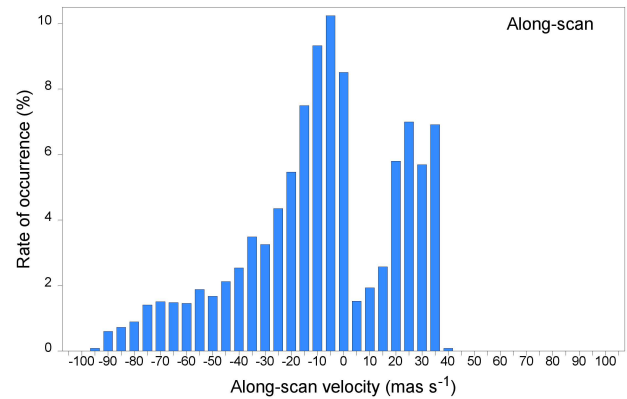}
\caption{\label{fig:figure4} The distribution of the apparent proper (relative to stars) along-scan velocity of Earth's Trojan asteroid, 2010~TK$_7$. The median absolute velocity is $\sim 40$~mas~s$^{-1}$, about six times greater than that of the Main Belt asteroids. }
\end{figure}

\begin{figure}
\includegraphics{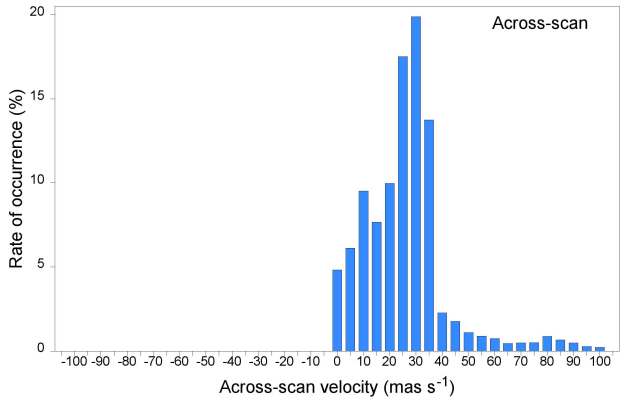}
\caption{\label{fig:figure5} The distribution of the across-scan velocity of Earth's Trojan asteroid, 2010~TK$_7$. The asymmetry is a result of 2010~TK$_7$ co-orbiting the Sun in Earth's L4 Lagrangian region. }
\end{figure}

Using the simulation data to consider the implications for observing a larger Earth Trojan in the same orbit we find that the along-scan (Fig.~\ref{fig:figure4}) and across-scan (Fig.~\ref{fig:figure5}) velocities are considerably greater than Main Belt asteroids. The median absolute along-scan velocity of 2010~TK$_7$  (Fig.~\ref{fig:figure4}) is $\sim 40$~mas~s$^{-1}$, almost six times that of the Main Belt asteroids. A motion along-scan of this magnitude results in the object drifting out of the 6~x~6 pixels window soon after crossing the first CCD in \textit{Gaia}'s focal plane array.

The across-scan motion of 2010~TK$_7$ (Fig.~\ref{fig:figure5}) is also large compared to that of Main Belt asteroids. The absolute across-scan velocity of Main Belt asteroids is typically less than 15~mas~s$^{-1}$ \citep{2012arXiv1212.0268T} compared with 40~mas~s$^{-1}$ for 2010~TK$_7$. We also note that the distribution is asymmetric and not the bimodal distribution we observed with the simulated Earth Trojans in \citet{2012arXiv1212.0268T}. We attribute this asymmetry to the peculiarity of 2010~TK$_7$ co-orbiting the Sun in Earth's L4 Lagrangian region.

Given the 106.5\degr separation between \textit{Gaia}'s lines of sight, motion in the across-scan direction of 400~mas~s$^{-1}$ would result in an object traversing the width of the focal plane array (0.7\degr) between the first and second observations of the field \citep{2012MNRAS.424..372T}. As the across-scan motion of 2010~TK$_7$ is typically less than 40~mas~s$^{-1}$ we find that a kilometre-scale Earth Trojan in the same orbit as 2010~TK$_7$ would provide enough signal due to its proximity and hence greater apparent brightness to be detected even though it would drift out of the 6~x~6 pixels window soon after crossing the first CCD in \textit{Gaia}'s focal plane array. In this orbit an object will drift approximately 4~arcmin in the interval between the first and second observations of the field thus making detections on both crossings possible.

The conclusion that can be drawn from this is that if a kilometre-scale Earth Trojan does exist then \textit{Gaia} will detect it in spite of the signal degradation caused by the high relative drift across the CCD array.
There will also be a substantial number of CCD crossings where the relative drift is low enough that \textit{Gaia} will detect any Earth Trojans with diameters larger than $\sim 600$~m.

\subsection{Mars Trojans}

As with the simulation for the Earth Trojan 2010~TK$_7$ we replicated the orbital parameters for the Mars Trojans (Table~\ref{tab:table2}) and simulated observations by \textit{Gaia}. The three largest Trojans (Eureka, 1998~VF$_{31}$ and 1999~UJ$_7$) were fully detected in our simulations. Of the remaining Trojans, 2007~NS$_2$ had a 98 per cent detection, 2001~DH$_{47}$ had a 43 per cent detection and 2011~SL$_{25}$ had a nine per cent detection. The Trojans 2011~SC$_{191}$ and 2011~UN$_{63}$ were not detected in our simulations.

The apparent magnitude of a Mars Trojan with a diameter of 1~km and an albedo of 0.20 ranges between $V = 16.2$ at opposition and $V = 20.7$ at a Solar elongation of $60\degr$ \citep{2012MNRAS.424..372T}. \textit{Gaia} cannot observe at elongations within $45\degr$ of opposition, however, the $\sim 1$~km Trojans are bright enough to be detected when they are in \textit{Gaia}'s field of view. 

The diameters of the Mars Trojans (Table~\ref{tab:table4}) were calculated using the standard formula \citep{2009Icar..202..134W}: 
\begin{equation}
\log D = 3.1235 - 0.2 H - 0.5 \log p_{v} ,
\end{equation}
where $D$ is diameter in kilometres, $H$ is absolute magnitude and $p_{v}$ is albedo. 

\citet{2003Icar..165..349R} found that Eureka has a visible spectrum consistent with the Sr class in the Bus taxonomy and an infrared spectrum that is consistent with the A class. They also found that 1998~VF$_{31}$ fit closely to Sr or Sa class and that 1999~UJ$_7$ fit to X or T class. \citet{2007Icar..192..442T} subsequently observed Eureka and 1998~VF$_{31}$ and determined their albedos to be 0.39 and 0.32 respectively. The other Trojans were assumed to be S class and to have typical albedo values as described in \citet{2009Icar..202..134W}.

\begin{table}
 \centering
\caption{Calculated diameters of Mars Trojans, assuming S class except where noted. \label{tab:table4}}
\begin{tabular}{lccr@{.}lc}
\hline 
Asteroid            & Abs. mag. & Class           & \multicolumn{2}{c}{Albedo}  & Diameter    \tabularnewline
                    & $(H)$     &                 & \multicolumn{2}{c}{ }       & (km)      \tabularnewline
\hline 
Eureka              & 16.1    & Sr$^{\dagger}$  & 0&39$^{\ddagger}$           & 1.28   \tabularnewline
1998~VF$_{31}$      & 17.1    & Sr$^{\dagger}$  & 0&32$^{\ddagger}$           & 0.89   \tabularnewline
1999~UJ$_7$         & 16.9    & X$^{\dagger}$   & 0&174$^{\#}$                & 1.33   \tabularnewline
2007~NS$_2$         & 17.8    & S               & 0&203$^{\#}$                & 0.81   \tabularnewline
2001~DH$_{47}$      & 18.7    & S               & 0&203$^{\#}$                & 0.54   \tabularnewline
2011~SC$_{191}$     & 19.3    & S               & 0&203$^{\#}$                & 0.41   \tabularnewline
2011~SL$_{25}$      & 19.5    & S               & 0&203$^{\#}$                & 0.37   \tabularnewline
2011~UN$_{63}$      & 19.7    & S               & 0&203$^{\#}$                & 0.34   \tabularnewline
\hline 
\multicolumn{6}{l}{$\dagger$ Spectral class taken from \citet{2003Icar..165..349R}} \tabularnewline
\multicolumn{6}{l}{$\ddagger$ Albedo taken from \citet{2007Icar..192..442T}} \tabularnewline
\multicolumn{6}{l}{$^{\#}$ Typical albedo values taken from \citet{2009Icar..202..134W}}  \tabularnewline
\end{tabular}
\end{table}

\begin{figure}
\includegraphics{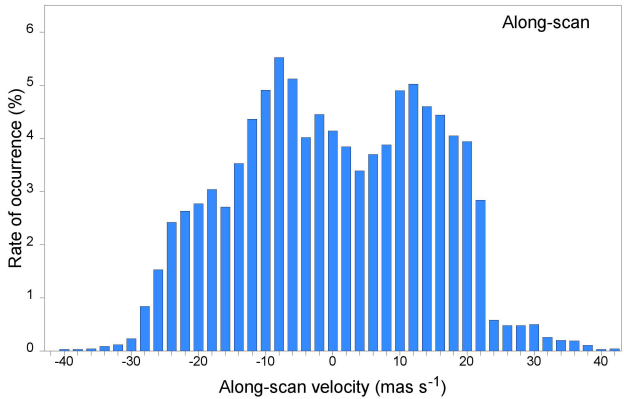}
\caption{\label{fig:figure6} The distribution of the apparent proper (relative to stars) along-scan velocity of the Mars Trojan asteroids. The median absolute velocity is $\sim 16$~mas~s$^{-1}$, about double that of the Main Belt asteroids.}
\end{figure}

\begin{figure}
\includegraphics{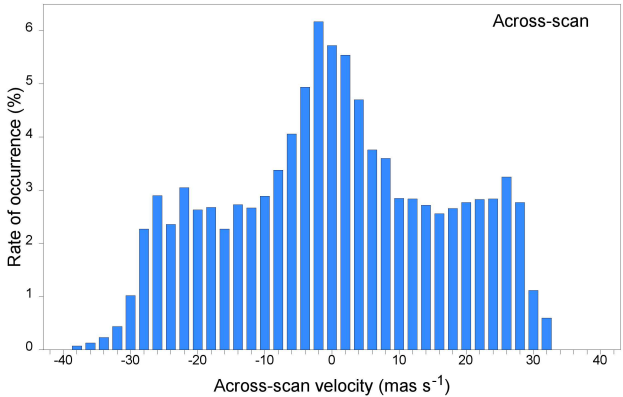}
\caption{\label{fig:figure7} The distribution of the across-scan velocity of Mars Trojan asteroids. The distribution is similar to the across-scan distribution for Main Belt asteroids.}
\end{figure}

As with the Earth Trojan 2010~TK$_7$ we found that the along-scan (Fig.~\ref{fig:figure6}) and across-scan (Fig.~\ref{fig:figure7}) velocities are greater than Main Belt asteroids. This result was not unexpected since the Mars Trojans are nearer to \textit{Gaia} than the Main Belt asteroids. Likewise, the velocities were greater for 2010~TK$_7$ than for the Mars Trojans. The median along-scan velocity of the Mars Trojans (Figure \ref{fig:figure6}) is $\sim 16$~mas~s$^{-1}$, about double that of the Main Belt asteroids. A motion along-scan of this magnitude results in the object drifting out of the 6~x~6 pixels window after crossing two CCDs in \textit{Gaia}'s focal plane array, resulting in a loss of signal.

The across-scan motion of the Mars Trojans (Fig.~\ref{fig:figure7}) is also large compared to that of Main Belt asteroids but not as large as that of 2010~TK$_7$. The absolute across-scan velocity of the Mars Trojans is typically less than 30~mas~s$^{-1}$, about double that of Main Belt asteroids. We note that the distribution bears some resemblance to the across-scan distribution for Main Belt asteroids in \citet{2012P&SS...73....5T}.

We find some differences between these simulation results and the expected results for Main Belt asteroids. It is expected that \textit{Gaia} will make 60 to 70 observations of each Main Belt asteroid over the five-year period of its mission \citep{2012P&SS...73....5T}. By comparison the number of observations of the Mars Trojans, except for Eureka, are somewhat fewer. We expect that the number of observations will be linked to the apparent magnitude, which will be a function of distance and phase angle. \citet{2012P&SS...73....5T} found a trend in the relationship between the absolute magnitude of an object and its apparent magnitude when observed by \textit{Gaia}. It was calculated that \textit{Gaia}'s limit for Main Belt asteroids is $H \simeq 16 \pm 1$ . Based on our simulation results we have determined that \textit{Gaia}'s limit for Mars Trojans is $H \simeq 19 \pm 0.5$.

An examination of the positions of the detections relative to Earth (and \textit{Gaia}) shows a link between the brightness and the number of detections. It follows that there is a link between the brightness and the range of elongations over which the Trojans can be detected (Table~\ref{tab:table5}). A result which was initially surprising was the variation in the along-scan and across-scan velocities, however, examination of the results shows that this variation is related to the orbit parameters.

\begin{table}
 \centering
\caption{Detection statistics for Mars Trojans. Distance from Earth represents the distance to the object at the nearest observation. Solar elongation represents the minimum angle between the Sun and the object at which an observation was made. Magnitude is the brightest magnitude at which an observation was made. Values for 2011~SC$_{191}$ and 2011~UN$_{63}$ are enclosed in parentheses as these are below \textit{Gaia}'s detection threshold. \label{tab:table5}}
\begin{tabular}{lccc}
\hline 
Asteroid                 & Distance from  & Solar    & Magnitude     \tabularnewline
                         & Earth $(AU)$   & elongation    & $(V)$    \tabularnewline
\hline 
Eureka            & 0.546        & 48\fdg 9      & 16.9       \tabularnewline
1998~VF$_{31}$  & 0.459        & 72\fdg 7      & 17.4       \tabularnewline
1999~UJ$_7$     & 0.557        & 74\fdg 5      & 17.8       \tabularnewline
2007~NS$_2$     & 0.548        & 100\fdg 2     & 18.6       \tabularnewline
2001~DH$_{47}$           & 0.557        & 124\fdg 3     & 19.6       \tabularnewline
2011~SC$_{191}$          & $(0.536)$    &               & $(20.1)$   \tabularnewline
2011~SL$_{25}$           & 0.448        & 128\fdg 2     & 19.8       \tabularnewline
2011~UN$_{63}$           & $(0.500)$    &               & $(20.3)$   \tabularnewline
\hline 
\end{tabular}
\end{table}

A comparison of the orbits (Table~\ref{tab:table2}) shows that the eccentricity of 1998~VF$_{31}$ and 2011~SL$_{25}$ is $e\simeq{0.1}$ and that the other Trojans are $e\simeq{0.05}$. It follows that the nearest approach to Earth's orbit and hence the distance from \textit{Gaia} for the nearest detection shows a similar pattern. It can be seen that the distances for 1998~VF$_{31}$ and 2011~SL$_{25}$ are similar to each other, as are the other Trojans (Table~\ref{tab:table5}). We see in the distribution of the along-scan velocity (Fig.~\ref{fig:figure6}) that most of the absolute velocities are less than 25~mas~s$^{-1}$. Absolute velocities greater than 25~mas~s$^{-1}$ are due to 1998~VF$_{31}$ and 2011~SL$_{25}$ approaching more closely to \textit{Gaia} as a result of the eccentricity of their orbits.

\begin{figure}
\includegraphics{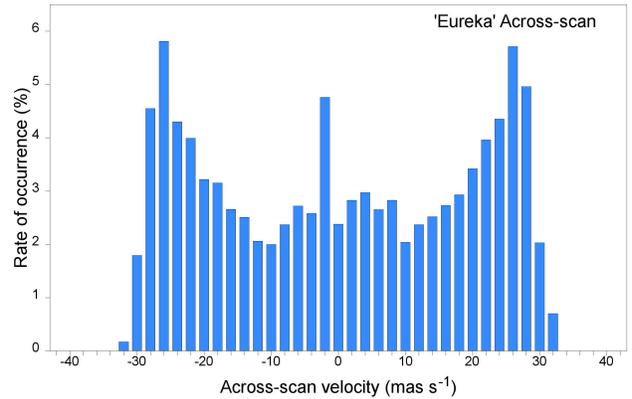}
\caption{\label{fig:figure8} The distribution of the across-scan velocity of Mars Trojan asteroid Eureka.}
\end{figure}

The case is not the same for the across-scan velocity distribution (Fig.~\ref{fig:figure7}) although most of the velocities are also less than 25~mas~s$^{-1}$. The distribution of across-scan velocities for most of the Mars Trojans exhibit this profile of a characteristic central peak with a shoulder either side. The maximum and minimum velocity for each Trojan is also consistent with the maximum and minimum velocity in the along-scan direction. However, the across-scan velocity distribution for Eureka is markedly different (Fig.~\ref{fig:figure8}) in that we see the across-scan velocities biased towards the maximum and minimum. We postulate that this is due to Eureka being bright enough to be detected at almost all Solar elongations and that these detections at the velocity extremes decrease with size and corresponding increase in minimum Solar elongation.

\begin{figure}
\includegraphics{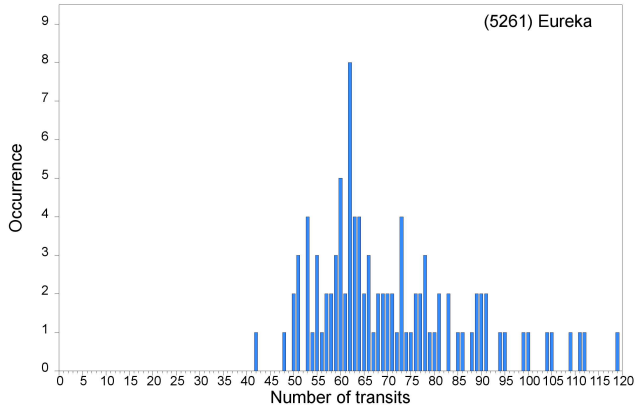}
\caption{\label{fig:figure9}The number of detections of Mars Trojan asteroid Eureka. A maximum occurs at 62 detections. }
\end{figure}

In our simulations of Eureka we found that it was fully detected, i.e. all the clones were detected multiple times. Depending on the starting conditions the number of detections of Eureka ranges between 42 and 141 detections with a probability peak at 62 detections (Fig.~\ref{fig:figure9}). This number of detections is similar to that expected for Main Belt asteroids. 

\begin{figure}
\includegraphics{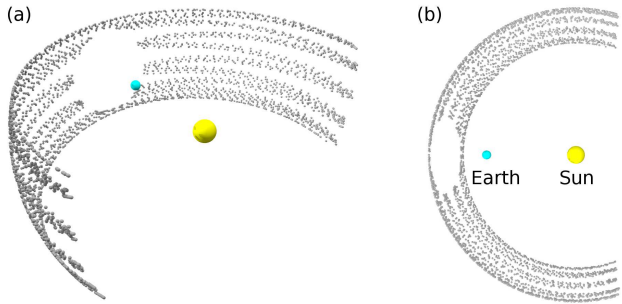}
\caption{\label{fig:figure10}(a) Plotting the detections of Mars Trojan asteroid Eureka relative to \textit{Gaia}'s orientation shows the range of Solar elongations at which observations can be made. The height of the region is a result of the orbit inclination. (b) The exclusion of observations at Solar elongations closer than $45\degr$ to the Sun results in a complementary exclusion zone within $45\degr$ of opposition. The apparent thickness of the region is an effect of the projection. }
\end{figure}

Plotting the detections of Eureka relative to Earth and \textit{Gaia} (Fig.~\ref{fig:figure10}) shows that observations are made at all Solar elongations between about ${49\degr}$ and $135\degr$. At its nearest approach at a Solar elongation of $135\degr$ Eureka is $\sim{0.5}$~AU from Earth and has a brightness of $V\simeq{16.9}$. Its farthest detectable distance is at a Solar elongation of $48\fdg 9$ where its brightness is at \textit{Gaia}'s detection threshold of $V\simeq{20}$.

\begin{figure}
\includegraphics{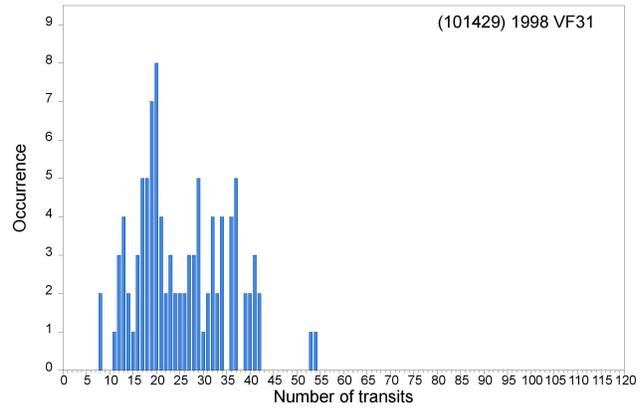}
\caption{\label{fig:figure11}The number of detections of Mars Trojan asteroid 1998~VF$_{31}$. A maximum occurs at 20 detections. }
\end{figure}

\begin{figure}
\includegraphics{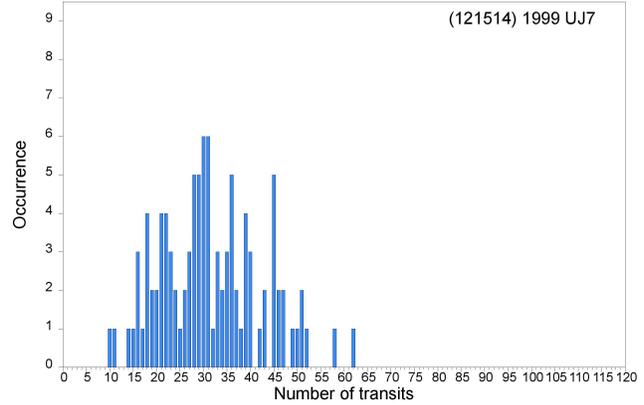}
\caption{\label{fig:figure12}The number of detections of Mars Trojan asteroid 1999~UJ$_7$. A maximum occurs at 30 detections. }
\end{figure}

We also found that 1998~VF$_{31}$ and 1999~UJ$_7$ were fully detected. The results for 1998~VF$_{31}$ and 1999~UJ$_7$ were quite similar to each other. We found that detections of 1998~VF$_{31}$ ranged between eight and 54 detections with a probability peak at 20 detections (Fig.~\ref{fig:figure11}) and for 1999~UJ$_7$ ranged between ten and 62 detections with a peak at 30 detections (Fig.~\ref{fig:figure12}).

\begin{figure}
\includegraphics{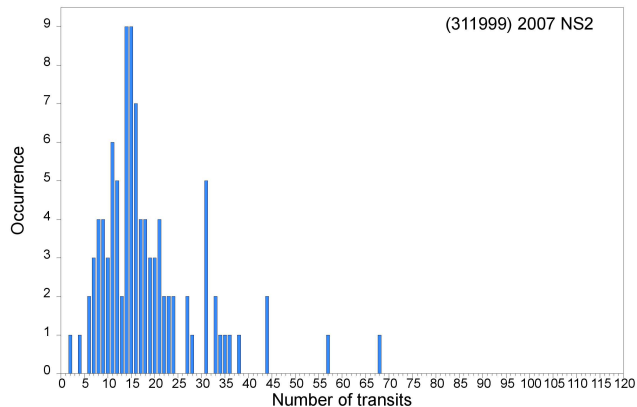}
\caption{\label{fig:figure13}The number of detections of Mars Trojan asteroid 2007~NS$_2$. A maximum occurs at 15 detections. }
\end{figure}

In our simulations of the remaining Mars Trojans we found that they were not fully detected. We had a 98 per cent success rate with 2007~NS$_2$ as two of the clones went undetected. The other 98 clones were detected up to 68 times with a probability peak at 15 detections (Fig.~\ref{fig:figure13}). For 2001~DH$_{47}$ only 43 per cent of the clones were detected. The maximum number of detections was 21 and the probability peak was very low at four detections. For 2011~SL$_{25}$ only nine per cent of the clones were detected and the maximum number of detections for a single clone was 14 detections. The Trojans 2011~SC$_{191}$ and 2011~UN$_{63}$ were not detected in our simulations as their magnitudes were $V > 20$ at their brightest.

\begin{figure}
\includegraphics{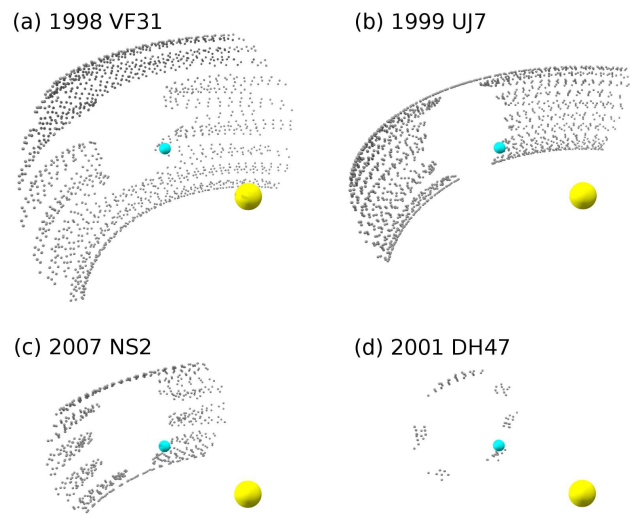}
\caption{\label{fig:figure14} The detections of Mars Trojan asteroids (a) 1998~VF$_{31}$, (b) 1999~UJ$_7$, (c) 2007~NS$_2$ and (d) 2001~DH$_{47}$ relative to \textit{Gaia}'s orientation. The heights of the regions are a result of the orbit inclinations. }
\end{figure}

Plotting the detections of 1998~VF$_{31}$, 1999~UJ$_7$, 2007~NS$_2$ and 2001~DH$_{47}$ relative to Earth and \textit{Gaia} (Fig.~\ref{fig:figure14}) shows the increasing minimum Solar elongations (respectively) at which observations were made (Table~\ref{tab:table5}). It is clear that the range of Solar elongations over which \textit{Gaia} will detect a Mars Trojan is related to its size and albedo and hence its range of apparent magnitudes. Eureka is the brightest of the Mars Trojans and has the largest range of Solar elongations at which detections can be made (Fig.~\ref{fig:figure10}). The other Mars Trojans are fainter and have smaller regions of detection (Fig.~\ref{fig:figure14}). The Trojans 1998~VF$_{31}$ (Fig.~\ref{fig:figure14}a) and 1999~UJ$_7$ (Fig.~\ref{fig:figure14}b) are similar in brightness and have similar ranges of Solar elongations where they can be detected. With 2007~NS$_2$ (Fig.~\ref{fig:figure14}c) and 2001~DH$_{47}$ (Fig.~\ref{fig:figure14}d) we see a reduced range of Solar elongations resulting in the detection region being an annulus around the exclusion zone towards opposition. The implication of this is that \textit{Gaia} will only detect smaller Trojans if they pass through this annulus.

From our simulations we find that the starting conditions and orbit parameters greatly influence the result. We expect that \textit{Gaia} will detect any kilometre-scale Mars Trojans. We can state with a high degree of certainty that \textit{Gaia} will detect Mars Trojans larger than 800~m in diameter. We also consider that \textit{Gaia} will detect Mars Trojans as small as 400~m in diameter. Consequently we expect that the results of the \textit{Gaia} mission will test the predictions about the Mars Trojan asteroid population and lead to greater understanding about the evolution of the Solar System.

\section{Summary}

We used the known orbital elements for the Earth and Mars Trojans to simulate detections by the \textit{Gaia} satellite. We cloned the orbital elements to provide 100 test bodies for each of the Trojans and modified them to compensate for the uncertainty of when \textit{Gaia} will begin observations. We found that \textit{Gaia} will not detect the Earth Trojan 2010~TK$_7$ but consider that it will detect any Earth Trojan with a diameter larger than 600~m. We also found that \textit{Gaia} will detect many of the Mars Trojans and is therefore likely to discover additional Trojans.

\subsection{Earth Trojans}

Although 2010~TK$_7$ is very close to Earth and for the most part is within the observable region of sky for \textit{Gaia} (Fig.~\ref{fig:figure3}), our simulation results indicate that it will not be detected by \textit{Gaia}. Due to its small size of $\sim 300$~m \citep{2011Natur.475..481C} its brightness is low, ranging from $V = 20.9$ to $V = 22.8$ (Table~\ref{tab:table3}). Consequently 2010~TK$_7$ is always too faint to be detected by \textit{Gaia}.

We used the simulation results to consider the implications for observing a larger Earth Trojan in the same orbit. The along-scan (Fig.~\ref{fig:figure4}) and across-scan (Fig.~\ref{fig:figure5}) velocities are considerably greater than Main Belt asteroids. The median absolute along-scan velocity is $\sim 40$~mas~s$^{-1}$, almost six times that of the Main Belt asteroids, and results in the object drifting out of the 6~x~6 pixels window after crossing just one of the CCDs in \textit{Gaia}'s focal plane array. The across-scan motion is typically less than $40$~mas~s$^{-1}$, about three times higher than Main Belt asteroids. We note that the distribution of the across-scan motion is asymmetric and not the bimodal distribution we found with the simulated Earth Trojans in \citet{2012arXiv1212.0268T}. We attribute this asymmetry to the object residing in Earth's L4 Lagrangian region.

We conclude that, if a kilometre-scale Earth Trojan does exist, then \textit{Gaia} will detect it in spite of the signal degradation caused by the relative drift across the CCD array. We also consider that \textit{Gaia} will detect any Earth Trojan with a diameter larger than $\sim 600$~m as there will be a substantial number of CCD crossings where the relative drift is low enough that the signal degradation will be minimal.

\subsection{Mars Trojans}

As with the simulation for the Earth Trojan 2010~TK$_7$ we replicated the orbital parameters for the Mars Trojans (Table~\ref{tab:table2}). The three largest Trojans (Eureka, 1998~VF$_{31}$ and 1999~UJ$_7$) were fully detected in our simulations. Of the remaining Trojans, 2007~NS$_2$ had a 98 per cent detection, 2001~DH$_{47}$ had a 43 per cent detection and 2011~SL$_{25}$ had a nine per cent detection. The Trojans 2011~SC$_{191}$ and 2011~UN$_{63}$ were not detected in our simulations.

As with the Earth Trojan we found that the along-scan (Fig.~\ref{fig:figure6}) and across-scan (Fig.~\ref{fig:figure7}) velocities are greater than Main Belt asteroids. The median along-scan velocity of the Mars Trojans is $\sim 16$~mas~s$^{-1}$, about double that of the Main Belt asteroids. A motion along-scan of this magnitude results in the object drifting out of the 6~x~6 pixels window after crossing only two of the CCDs in \textit{Gaia}'s focal plane array. The across-scan motion of the Mars Trojans is typically less than 30~mas~s$^{-1}$, also about double that of Main Belt asteroids. We note that the distribution bears some resemblance to the across-scan distribution for Main Belt asteroids in \citet{2012P&SS...73....5T}.

We found that Eureka, 1998~VF$_{31}$ and 1999~UJ$_7$ were fully detected, i.e. all of the clones were detected. The number of observations of the Mars Trojans, except for Eureka, will be somewhat fewer than the expected 60 to 70 observations of each Main Belt asteroid \citep{2012P&SS...73....5T} over the five-year period of its mission. We found that the number of detections of Eureka is similar to that expected for Main Belt asteroids. Examining the positions of the detections relative to Earth (and \textit{Gaia}) shows that there is a link between the maximum brightness and hence the range of elongations over which the Trojans can be detected (Table~\ref{tab:table5}). 

The results for 1998~VF$_{31}$ and 1999~UJ$_7$ were quite similar, with probability peaks at 20 and 30 detections respectively. However, the remaining Mars Trojans were not fully detected. For 2007~NS$_2$ and 2001~DH$_{47}$ we found the probability of detection to be 98 and 43 per cent respectively, with corresponding probability peaks at 15 and four detections. For 2011~SL$_{25}$ only nine per cent of the clones were detected. The Trojans 2011~SC$_{191}$ and 2011~UN$_{63}$ were not detected in our simulations as their magnitudes were $V > 20$ at their brightest.

From our simulations we find that the starting conditions and orbit parameters greatly influence the result in the case of detecting smaller Mars Trojans. Based on our simulation results we have determined that the \textit{Gaia} limit for the detection of Mars Trojans is $H \simeq 19 \pm 0.5$. We expect that \textit{Gaia} will detect any kilometre-scale Mars Trojans. We can also state with a high degree of certainty that \textit{Gaia} will detect Mars Trojans larger than 800~m in diameter and that \textit{Gaia} may detect Mars Trojans as small as 400~m in diameter. 

Past simulations \citep{1999ApJ...517L..63T, 2000MNRAS.319...63T, 2000MNRAS.319...80E} predict that the number of Mars Trojans with diameters larger than 1~km could be more than 50 and that the number with diameters larger than 100~m could be several hundred. Assuming a progression of sizes that follows a power law, the number of Mars Trojans that \textit{Gaia} will discover with diameters larger than 800~m will be more than 85, or conservatively more than 55. If we take the conservative view that there are about 30 Mars Trojans which have diameters larger than 1~km then we should expect that \textit{Gaia} will detect between 55 and 140 new Mars Trojans. The results of the \textit{Gaia} mission will test not only the accuracy of these simulation results but also the existing models of the Mars Trojan population and lead to greater understanding about the evolution of the Solar System.

\section*{Acknowledgments}
MT acknowledges support from the Astronomical Society of Australia, the Australian Institute of Physics and the sponsoring organisations of the Gaia-FUN-SSO2 workshop. The work reported on in this publication has been supported by the European Science Foundation (ESF), in the framework of the GREAT Research Networking Programme. MT thanks the SOC/LOC of the Gaia-FUN-SSO2 workshop for providing a fertile environment for discussing \textit{Gaia} science. MT thanks Heidi Ursula Wallon Pizarro for feedback which helped to improve the manuscript. DMC is supported by an Australian Research Council Future Fellowship.

\label{lastpage}

\end{document}